\documentclass[conference]{IEEEtran}
\makeatletter
\def\ps@headings{%
\def\@oddhead{\mbox{}\scriptsize\rightmark \hfil \thepage}%
\def\@evenhead{\scriptsize\thepage \hfil \leftmark\mbox{}}%
\def\@oddfoot{}%
\def\@evenfoot{}}
\makeatother
\usepackage{cite}
\usepackage{citesort}
\usepackage{graphicx}
\usepackage{epstopdf}
\usepackage{stfloats}
\usepackage{amsmath}
\usepackage{enumitem}
\usepackage{amsfonts}
\usepackage{nicefrac}
\usepackage{amssymb,bm,upgreek}
\usepackage{algorithm}
\usepackage{nicefrac}
\usepackage{algpseudocode}

\begin{document}

\title{Mitigating Pilot Contamination Through Location-Aware Pilot Assignment in Massive MIMO Networks}

\author{\IEEEauthorblockN{Noman Akbar, Shihao Yan, Nan Yang, and Jinhong Yuan$^\dagger$}
\IEEEauthorblockA{Research School of Engineering, Australian National University, Acton, ACT 2601,
Australia\\$^\dagger$School of Electrical Engineering and Telecommunications, University of New South Wales, Sydney, Australia}Email: \{noman.akbar, shihao.yan, nan.yang\}@anu.edu.au, j.yuan@unsw.edu.au}

\maketitle

\begin{abstract}
We propose a novel location-aware pilot assignment scheme to mitigate pilot contamination in massive  multiple-input multiple-output (MIMO) networks, where the channels are subjected to Rician fading. Our proposed scheme utilizes the location information of users as the input to conduct pilot assignment in the network. Based on the location information, we first determine the line of sight (LOS) interference between the intended signal and the interfering signal. Our analysis reveals that the LOS interference converges to zero as the number of antennas at the base station (BS) goes to infinity, whereas for finite number of antennas at the BS the LOS interference indeed depends on specific pilot allocation strategies. Following this revelation, we assign pilot sequences to all the users in the massive MIMO network such that the LOS interference is minimized for finite number of antennas at the BS. Our proposed scheme outperforms the random pilot assignment in terms of achieving a much higher uplink sum rate for reasonable values of the Rician K-factor. Moreover, we propose a new performance metric, which measures the strength of the LOS interference and demonstrate that it is a good metric to analyze the performance of pilot assignment schemes. Theoretical analysis is provided and verified through numerical simulations to support the proposed scheme.
\end{abstract}

\section{Introduction}
Fifth generation (5G) mobile networks will offer a significantly higher throughput than the fourth generation (4G) mobile networks. Massive multiple-input multiple-output (MIMO) is considered as the key enabling technology to realize the promise of the increased throughput. Recently, an extensive research conducted by the mobile and wireless communications enablers for the twenty-twenty information society (METIS) strengthens the role of massive MIMO in 5G \cite{Aziz2015}. A significant finding of METIS is that implementing massive MIMO provides a twenty times increased throughput than the 4G networks, which is by far the largest contributing factor. Of course, we need to overcome certain challenges such as pilot contamination to fully unlock the lucrative advantages offered by massive MIMO.

Previous studies shows that pilot contamination severely limits the performance of massive MIMO \cite{Fernandes2013,Jose2011}. In principle, pilot contamination occurs when there are not enough orthogonal pilot sequences available in the massive MIMO network. Specifically, pilot contamination exists when the number of users in the network exceeds the length of the pilot sequence. The adverse nature of pilot contamination is highlighted by the fact that it does not disappear when the base station (BS) employs a very large number of antennas. Consequently, an increasing attention has been paid to alleviate the detrimental impact of pilot contamination in massive MIMO. The previous research on the topic falls into one of the five categories: 1) protocol based methods \cite{Fernandes2013}; 2) precoding based methods \cite{Jose2011}; 3) angle-of-arrival (AoA) based methods \cite{Yin2013}; 4) blind methods \cite{Muller2014}; and 5) pilot sequence design methods \cite{Akbar2016,Shen2015}. A new direction of research in mitigating pilot contamination is to leverage the known user location information and perform location-based processing, e.g., location-aware pilot assignment. In telecommunication networks, user location information can be easily obtained by either estimating the user location \cite{Abu-Shaban2016} or by obtaining the true user location from the global position system \cite{Taranto2014,Yan2016}. Location-aware pilot assignment schemes are generally focused on assigning pilot sequences to all the users in the network such that pilot contamination is minimized \cite{Muppirisetty2015,Zhao2015,Wang2015}. We note that the existing location-aware pilot assignment schemes \cite{Muppirisetty2015,Zhao2015,Wang2015} assume a simple channel model, which may not be generalized enough to depict certain channel conditions. Additionally, the existing works utilize location information to make the AoA of interfering signal non-overlapping \cite{Muppirisetty2015,Zhao2015,Wang2015}. Different from previous works, we make use of the deterministic location-dependent line of sight (LOS) channel component to assign pilot sequences in the network.

In this paper, we propose a location-aware pilot assignment scheme to improve the uplink sum rate performance by means of reducing pilot contamination in massive MIMO network. Different from \cite{Muppirisetty2015,Zhao2015,Wang2015}, we assume a Rician fading channel model in our analysis. We specifically focus on the LOS component of the received uplink signal and develop a location-aware algorithm to minimize the interference caused by the LOS component. We observe that minimizing LOS interference also results in reduced pilot contamination. Our proposed location-aware algorithm returns a user assignment matrix, which identifies the users that should be assigned the same pilot sequences to improve the uplink sum rate. The main contributions of our paper are summarized as follows:
\begin{enumerate}
  \item We investigate the LOS interference in massive MIMO networks. We find that the LOS interference is minimum for a certain pilot assignment in the network. Based on this finding, we propose a pilot assignment scheme that aims at mitigating the pilot contamination in massive MIMO networks.
  \item We note that the LOS interference converges to zero in massive MIMO regime. However, we find that the rate at which it converges to zero depends on the pilot assignment in the network. As such, our proposed pilot assignment scheme exploits the LOS interference convergence rate to assign same pilot sequences to the users that will result in reduced pilot contamination.
  \item We propose a new performance metric based on the LOS interference. The usefulness of the proposed metric is supported by the fact that uplink sum rate increases when the value of the proposed performance metric decreases.
  \item We compare the proposed location-aware pilot assignment scheme with the random pilot assignment scheme to demonstrate its advantages. We show that our proposed pilot assignment scheme minimizes the LOS interference and achieves a much higher uplink sum-rate than the random pilot assignment scheme.
\end{enumerate}

\section{System Model}
\begin{figure}
  \centering
  \includegraphics[width=15.5pc]{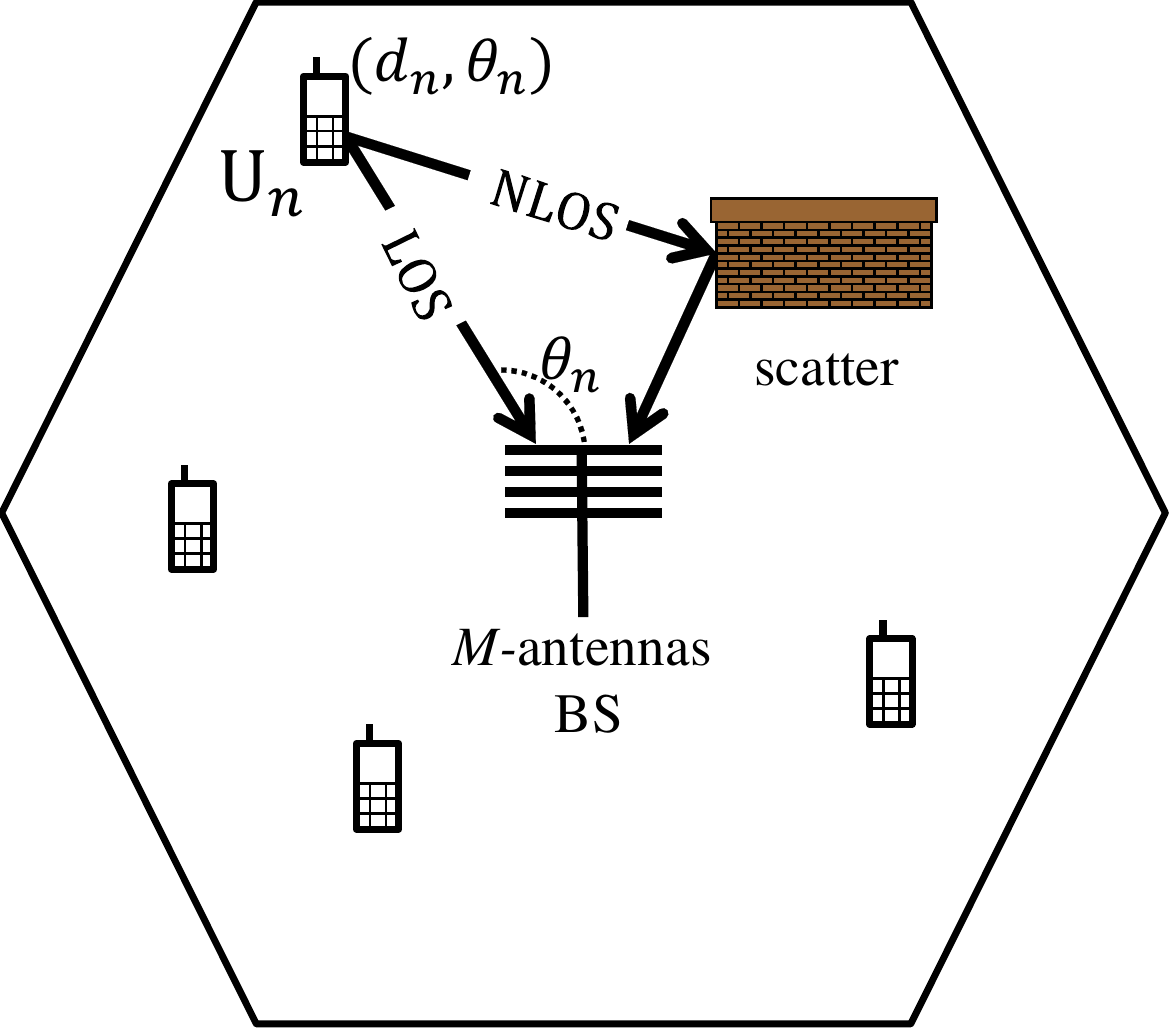}\\
  \caption{Illustration of the massive MIMO network of interest. The uplink channel between the user and the BS is subject to Rician fading, which consists of a LOS component and an NLOS component.}\label{sys_model}
\end{figure}

We consider a single-cell massive MIMO network consisting of a BS equipped with an \textit{M}-antenna uniform linear array and $N$ single-antenna users as depicted in Fig.~\ref{sys_model}. We denote the uplink channel between the \textit{n}-th user, i.e., $\textrm{U}_n$ and the \textit{m}-th BS antenna as ${g}_{mn} = \sqrt{\beta_{n}}{h}_{mn}$, where $\beta_{n}$ and ${h}_{mn}$ represents the large-scale and the small-scale fading coefficients, respectively. Furthermore, we assume that the channel ${g}_{mn}$ is subject to Rician fading, where the small-scale fading coefficient ${h}_{mn}$ consist of a deterministic LOS component and a non line of sight (NLOS) component denoted as ${h}_{mn}^{\textrm{LOS}}$ and ${h}_{mn}^{\textrm{NLOS}}$, respectively. The location of $\textrm{U}_n$ can be fully described by the tuple $\left(d_n,\theta_n\right)$, where $d_n$ and $\theta_n$ is the distance of $\textrm{U}_n$ from the BS and the AoA of $\textrm{U}_n$ at the BS, respectively. Furthermore, we assume that the Rician \textit{K}-factors for all the user are also know to the BS, which is a widely adopted assumption in the literature \cite{Zhang2014,Yan2016}. The LOS component of the small-scale fading coefficient of $\textrm{U}_n$ depends on $\theta_n$ and is given as \cite{Zhang2014}
\begin{align}\label{los_compo}
{h}_{mn}^{\textrm{LOS}} =& e^{-j\left(m-1\right)\left(\frac{2\pi d}{\lambda}\right)\textrm{sin}\left(\theta_n\right)},
\end{align}
where $d$ is the separation between two antenna elements of the uniform linear antenna array, $\lambda$ represents the wavelength of the carrier frequency, and $\theta_n$ denotes the AoA of $\textrm{U}_n$. Under the Rician fading model, the $M\times 1$ channel between $\textrm{U}_n$ and the BS is represented as
\begin{align}\label{full_chanl}
    \mathbf{g}_{n} =  \sqrt{\beta_n}\left(\sqrt{\frac{K_n}{K_n+1}}\mathbf{h}_{n}^{\textrm{LOS}} + \sqrt{\frac{1}{K_n+1}}\mathbf{h}_{n}^{\textrm{NLOS}}\right),
\end{align}
where $K_n$ denotes the Rician \textit{K}-factor for $\textrm{U}_n$, $\mathbf{g}_{n} = [\mathbf{g}_{1n},\mathbf{g}_{2n},\dotsc,\mathbf{g}_{Mn}]^{T}$, $\mathbf{h}_{n}^{\textrm{LOS}} = [{h}_{1n}^{\textrm{LOS}},{h}_{2n}^{\textrm{LOS}},\dotsc,{h}_{Mn}^{\textrm{LOS}}]^{T}$, and $\mathbf{h}_{n}^{\textrm{NLOS}} = [{h}_{1n}^{\textrm{NLOS}},{h}_{2n}^{\textrm{NLOS}},\dotsc,{h}_{Mn}^{\textrm{NLOS}}]^{T}$. Under the assumption of Rayleigh distributed NLOS component, $\mathbf{h}_{n}^{\textrm{NLOS}}$ follows a Gaussian distribution with zero mean and unit variance, i.e., $h_{mn}^{\textrm{NLOS}} \sim\mathcal{CN}(0,1)$. Furthermore, we assume that $\beta_n$ depends on the distance from the BS and is given as $\beta_n = \nicefrac{1}{(r_n/r_h)^v}$ \cite{Zhang2014}, where $r_n$ is the distance of $\textrm{U}_n$ from the BS, $r_h$ is the radius of the cell, and $v$ denotes the path loss exponent. We highlight that the reference power at $r_h$ is one.

The communication between the BS and the users consists of uplink transmission and downlink transmission phase. In this paper, we only focus on the uplink transmission phase. The uplink transmission phase is further divided into uplink training and uplink data transmission phase.

\subsection{Uplink Training}
 During the uplink training phase, all users transmit their pre-assigned pilot sequences to the BS with transmit power $p_u$. We assume that appropriate pilot sequence assignment has already been performed before the uplink training phase. The length of each pilot sequence is denotes as $\tau$. The BS utilizes the known $\tau \times N$ pilot sequence matrix $\mathbf{\Phi}$ to perform channel estimation, where the \textit{n}-th column of $\mathbf{\Phi}$ denotes the pilot sequence assigned to $\textrm{U}_n$. Accordingly, the $M \times \tau$ uplink pilot transmission matrix received at the BS is given as
\begin{align}\label{uplink_tx}
  \mathbf{Y} &= \sqrt{p_p}\mathbf{G}\mathbf{\Phi}^T + \mathbf{W},
\end{align}
where $p_p=\tau p_u$, $\mathbf{G} = [\mathbf{g}_{1},\mathbf{g}_{2},\dotsc,\mathbf{g}_{n}]$, and $\mathbf{W}$ represents the $M\times \tau$ additive white gaussian noise (AWGN) at the BS. The \textit{n}-th column of $\mathbf{G}$ represent the uplink channel from $\textrm{U}_n$ to the BS. We assume that the LOS component of the channel is known at the BS \cite{Yan2016}. Accordingly, the part of the received pilot transmission corresponding to the NLOS component is found from \eqref{uplink_tx} as
\begin{align}\label{uplink_tx1}
  \mathbf{Y}_\textrm{NLOS} & = \mathbf{Y} - \sqrt{p_p}\mathbf{{G}_{\textrm{LOS}}}\mathbf{\Phi}^T, \notag \\
                              & = \sqrt{p_p}\mathbf{{G}_{\textrm{NLOS}}}\mathbf{\Phi}^T + \mathbf{W},
\end{align}
where we define $\mathbf{{G}_{\textrm{LOS}}} = [\mathbf{g}_{1}^{\textrm{LOS}},\mathbf{g}_{2}^{\textrm{LOS}},\dotsc,\mathbf{g}_{N}^{\textrm{LOS}}]$ and $\mathbf{{G}_{\textrm{NLOS}}} = [\mathbf{g}_{1}^{\textrm{NLOS}},\mathbf{g}_{2}^{\textrm{NLOS}},\dotsc,\mathbf{g}_{N}^{\textrm{NLOS}}]$. Based on \eqref{uplink_tx1}, the BS estimates the NLOS component of the channel matrix using least square (LS) channel estimation as
\begin{align}\label{chan_est}
  \mathbf{\hat{G}}_{\textrm{NLOS}} &= \mathbf{Y}_\textrm{NLOS}\frac{1}{\sqrt{p_p}}\mathbf{\Phi}, \notag \\
                                   &=\mathbf{{G}_{\textrm{NLOS}}}\mathbf{\Phi}^T\mathbf{\Phi} + \frac{1}{\sqrt{p_p}}\mathbf{W}\mathbf{\Phi}, \notag \\
                                   &= \mathbf{{G}_{\textrm{NLOS}}}\mathbf{R_\Phi} + \frac{1}{\sqrt{p_p}}\mathbf{N},
\end{align}
where $\mathbf{N} = \mathbf{W}\mathbf{\Phi}$ and $\mathbf{R_\Phi} = \mathbf{\Phi}^T\mathbf{\Phi}$ denotes an $N \times N$ symmetric pilot correlation matrix. 
 Therefore, the overall estimated channel is given as
\begin{align}\label{full_est_chan}
  \mathbf{\hat{G}} &=   \mathbf{G}_{\textrm{LOS}} + \mathbf{\hat{G}}_{\textrm{NLOS}}.
\end{align}

We note that when $\tau = N$, it is possible that the pilot correlation matrix $\mathbf{R_\Phi} = \mathbf{I}_N$. As such, the only error during the channel estimation comes form noise, which can be confirmed from \eqref{chan_est}. Hence, pilot contamination can be eliminated when $\tau = N$. In this work we assume that $\tau \leq N$. We denote $\Psi_{m}$ as the group of users that are assigned the \textit{m}-th pilot sequence, where $m=\left\{1,2,\dotsc,\tau\right\}$. Accordingly, the channel estimate for $\textrm{U}_n \in \Psi_{m}$ is obtained from \eqref{chan_est} as
\begin{align}\label{full_est_chan12}
  \mathbf{\hat{g}}_n &=  \mathbf{g}_n^{\textrm{LOS}} + \mathbf{g}_n^{\textrm{NLOS}} +\sum_{\substack{j\in \Psi_{m} \\ j\neq n}}\mathbf{g}_j^{\textrm{NLOS}} + \frac{1}{\sqrt{p_p}}\mathbf{N}_n ,
\end{align}
where $\mathbf{N}_n$ denotes the \textit{n}-th column of $\mathbf{N}$. The channel estimate in \eqref{full_est_chan12} is contaminated due to the reuse of pilot sequences in the network, which occurs when same pilot sequences are assigned to $\textrm{U}_n$ and $\textrm{U}_j$. This phenomenon is known as \textit{pilot contamination} and is considered as a major bottleneck in massive MIMO.

\subsection{Uplink Data Transmission and Sum Rate}
During the uplink data transmission phase, $\textrm{U}_n$ transmits the data symbol $x_n$ to the BS, where $\mathbb{E}[x_n^H x_n] = 1$. After linear reception, the $N \times 1$ received signal at the BS is given by
\begin{align}\label{rec_sig}
  \mathbf{\hat{r}} =& \mathbf{\hat{A}}^{H}\left(\sqrt{p_u}\mathbf{G}\mathbf{x} + \mathbf{n}\right),
\end{align}
where, $\mathbf{\hat{A}}$ denotes an $M \times N$ linear detector matrix and $\mathbf{n}$ is the AWGN at the BS. Using \eqref{rec_sig}, the received signal corresponding to $\textrm{U}_n$ is written as
\begin{align}\label{rec_sig_n}
  \hat{r}_n =&  \sqrt{p_u}\mathbf{\hat{a}}_n^H\mathbf{g}_n x_n + \sqrt{p_u}\sum_{i=1,i\neq n}^{N}\mathbf{\hat{a}}_n^H\mathbf{g}_i x_i + \mathbf{\hat{a}}_n^H \mathbf{n},
\end{align}
where, $\mathbf{\hat{a}}_n$ is the \textit{n}-th column of $\mathbf{\hat{A}}$. We highlight that the last two terms in \eqref{rec_sig_n} is effective noise. Consequently, the ergodic achievable uplink rate for $\textrm{U}_n$ is given as \cite{Ngo2013,Zhang2014}
\begin{align}\label{uplink_sinr_n}
  \textrm{R}_n =& \mathbb{E}\left\{\textrm{log}_2\left(1+\frac{p_u|\mathbf{\hat{a}}_n^H\mathbf{g}_n|^2}{p_u \sum_{i=1,i\neq n}^{N} |\mathbf{\hat{a}}_n^H\mathbf{g}_i|^2 + \|\mathbf{\hat{a}}_n\|^2} \right)\right\}.
\end{align}
The uplink sum rate of the network is defined as
\begin{align}\label{sum_rate}
  \textrm{R} =& \frac{T-\tau}{T}\sum_{i=1}^{N} \textrm{R}_i,
\end{align}
where $T$ is the coherence time interval of the channel.


\section{Location-Aware Pilot Assignment}
In the section, we propose a location-aware pilot assignment scheme to improve the uplink sum rate of the massive MIMO network. We assume that location information for all the users in the network is available at the BS. 
As such, $\mathbf{G}_{\textrm{LOS}}$ in \eqref{full_est_chan} can be found by utilizing the known parameters.
\subsection{LOS Interference in Massive MIMO}
We now focus on the uplink rate expression given in \eqref{uplink_sinr_n}. Assuming the BS performs zero-forcing (ZF) detection, the interference term in the denominator of \eqref{uplink_sinr_n} is rewritten as
\begin{align}\label{interference_func}
  \mathbf{\hat{a}}_n^H \mathbf{g}_i=& \left[\frac{\mathbf{\hat{g}}_n}{\mathbf{\hat{g}}_n^H \mathbf{\hat{g}}_n}\right]^H \mathbf{g}_i = \frac{\mathbf{\hat{g}}_n^H   \mathbf{g}_i }{\mathbf{\hat{g}}_n^H \mathbf{\hat{g}}_n}.
\end{align}
Using \eqref{full_chanl} and \eqref{full_est_chan12}, we calculate the numerator in \eqref{interference_func} as
\begin{align}\label{interference_func1}
  \mathbf{\hat{g}}_n^H \mathbf{g}_i =& \left(\mathbf{g}_n^{\textrm{LOS}}\right)^H\mathbf{g}_i^{\textrm{LOS}} + \left(\mathbf{g}_n^{\textrm{LOS}}\right)^H\mathbf{g}_i^{\textrm{NLOS}} + \left(\mathbf{g}_n^{\textrm{NLOS}}\right)^H\mathbf{g}_i^{\textrm{LOS}} \notag \\ +& \left(\mathbf{g}_n^{\textrm{NLOS}}\right)^H\mathbf{g}_i^{\textrm{NLOS}} + \sum_{\substack{j\in \Psi_{m} \\ j\neq n}}\left(\mathbf{g}_j^{\textrm{NLOS}}\right)^H \left(\mathbf{g}_i^{\textrm{LOS}} + \mathbf{g}_i^{\textrm{NLOS}}\right) \notag \\ +&\left(1/\sqrt{p_p}\mathbf{N}_n\right)^H\mathbf{g}_i^{\textrm{LOS}} + 1/\sqrt{p_p}\left(\mathbf{N}_n\right)^H\mathbf{g}_i^{\textrm{NLOS}}.
\end{align}
We note that \eqref{interference_func1} contains the multiplication of the LOS component of the channel vector for $\textrm{U}_n$ and the channel vector of $\textrm{U}_i$. 
We define the LOS interference term as $I_{ni} \triangleq (\mathbf{g}_n^{\textrm{LOS}})^H \mathbf{g}_i^{\textrm{LOS}} /{\mathbf{\hat{g}}_n^H \mathbf{\hat{g}}_n}$. Using $\eqref{los_compo}$, we obtain
\begin{align}\label{interference_func51}
(\mathbf{g}_n^{\textrm{LOS}})^H \mathbf{g}_i^{\textrm{LOS}} =&~\Omega_{ni}~\left[1+\dotsc+e^{j\left(M-1\right) d\theta_{ni}}\right],
\end{align}
where $d=\lambda/2$ is applied and we define $\Omega_{ni} = \sqrt{\nicefrac{\beta_n\beta_i K_n K_i}{\left(K_n+1\right)\left(K_i+1\right)}}$, $d\theta_{ni}  \triangleq \pi\left[\textrm{sin}\left(\theta_n\right) - \textrm{sin}\left(\theta_i\right)\right]$. Using the property of sum of exponentials, \eqref{interference_func51} is further simplified as
\begin{align}\label{interference_func2}
(\mathbf{g}_n^{\textrm{LOS}})^H \mathbf{g}_i^{\textrm{LOS}}=&~\Omega_{ni} \frac{e^{j M d\theta_{ni}} -1 }{e^{j d\theta_{ni}} -1}, \notag \\
                                  =&~\Omega_{ni}\frac{-e^{j\frac{M}{2} d\theta_{ni}}\left(-e^{-j\frac{M}{2} d\theta_{ni}}-e^{j\frac{M}{2} d\theta_{ni}}\right)}{-e^{j\frac{1}{2}d\theta_{ni}}\left(-e^{-j\frac{1}{2}d\theta_{ni}}-e^{j\frac{1}{2} d\theta_{ni}}\right)}, \notag \\
                                  =&~\Omega_{ni}\left[\frac{\textrm{sin}\left(\frac{M  d\theta_{ni}}{2}\right)}{\textrm{sin}\left(\frac{ d\theta_{ni}}{2}\right)}\right] e^{j d\theta_{ni}\left( \frac{M - 1}{2}\right) }.
\end{align}
Utilizing \eqref{full_est_chan12}, we now find the denominator in \eqref{interference_func} as
\begin{align}\label{interference_func_deno}
\frac{1}{M}{\mathbf{\hat{g}}_n^H \mathbf{\hat{g}}_n} =& \frac{1}{M}\left(\mathbf{g}_n^{\textrm{LOS}} + \mathbf{g}_n^{\textrm{NLOS}} +\sum_{\substack{j\in \Psi_{m} \\ j\neq n}}\mathbf{g}_j^{\textrm{NLOS}} + \frac{1}{\sqrt{p_p}}\mathbf{N}_n\right)^H \notag \\ \times &\left(\mathbf{g}_n^{\textrm{LOS}} + \mathbf{g}_n^{\textrm{NLOS}} +\sum_{\substack{j\in \Psi_{m} \\ j\neq n}}\mathbf{g}_j^{\textrm{NLOS}} + \frac{1}{\sqrt{p_p}}\mathbf{N}_n\right),
\end{align}
In massive MIMO regime, the channels become increasingly orthogonal such as
\begin{align}\label{ortho_mm}
\frac{1}{M}\mathbf{h}_{i}^{H}\mathbf{h}_{j}=
\begin{cases}
1, & \forall~i=j\\
0, & \text{otherwise.}
\end{cases}
\end{align}
Using this property, \eqref{interference_func_deno} is reduced to the following
\begin{align}\label{interference_func_deno_approx}
\frac{1}{M}{\mathbf{\hat{g}}_n^H \mathbf{\hat{g}}_n} \xrightarrow{a.s} \beta_n + \sum_{\substack{j\in \Psi_{m} \\ j\neq n}}\frac{\beta_j}{1+K_j} +\frac{1}{p_p}.
\end{align}
Now, from \eqref{interference_func2} and \eqref{interference_func_deno_approx} we find $|I_{ni}|^2$ as
\begin{align}\label{interference_func4}
  |I_{ni}|^2 =&  \frac{\Omega_{ni}^2}{M^2\left(\beta_n + \sum_{\substack{j\in \Psi_{m} \\ j\neq n}}\frac{\beta_j}{1+K_j} +\frac{1}{p_p}\right)^2} \left[\frac{\textrm{sin}\left(\frac{M d\theta_{ni}}{2}\right)}{\textrm{sin}\left(\frac{d\theta_{ni}}{2}\right)}\right]^2.
\end{align}

\begin{figure}
  \centering
  \includegraphics[width=18pc]{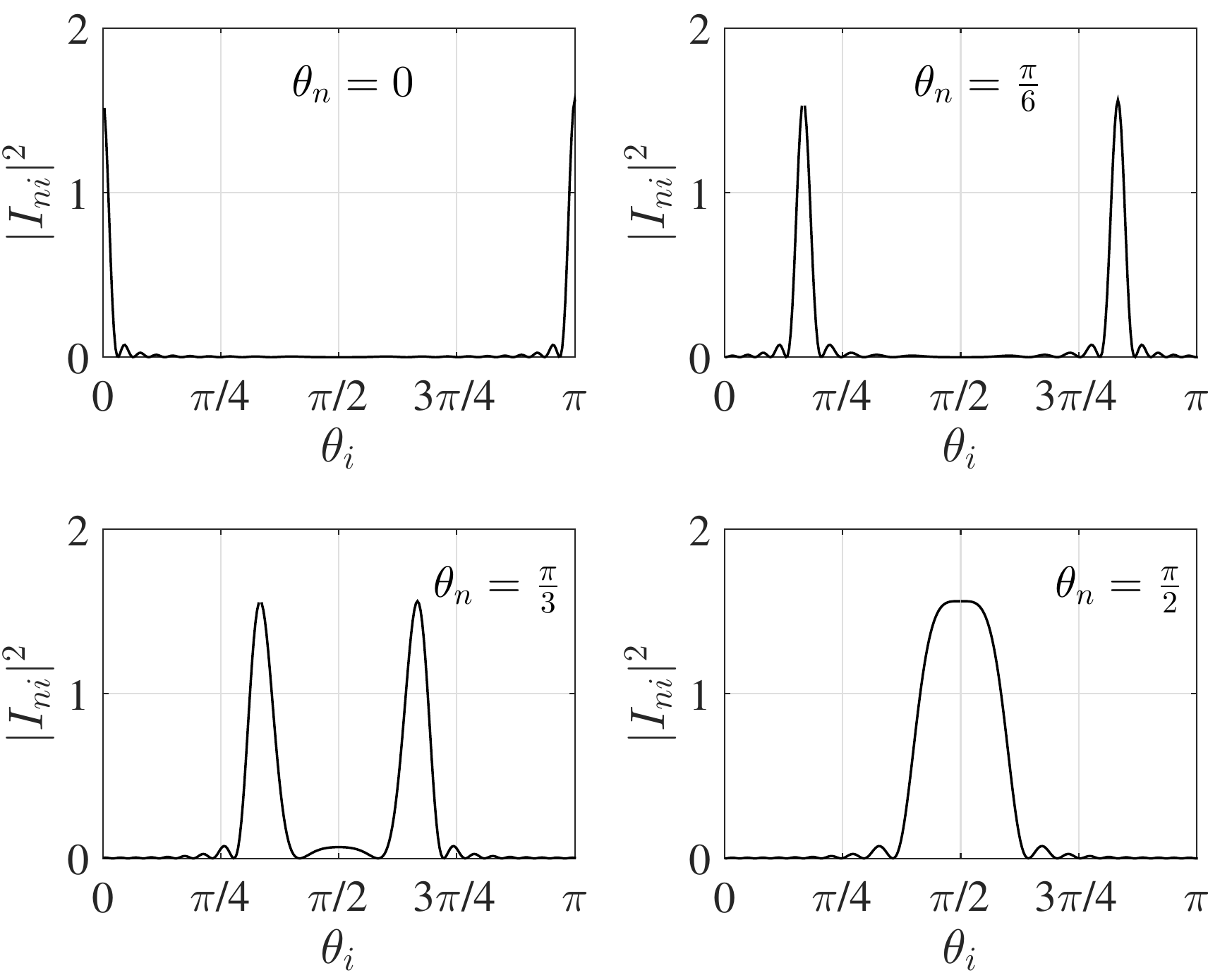}\\
  \caption{AoA versus the LOS interference between $\textrm{U}_n$ and $\textrm{U}_i$}\label{int_sin}
\end{figure}

We highlight that $|I_{ni}|^2$ gives a measure of interference between the LOS component of the uplink channels for $\textrm{U}_n$ and $\textrm{U}_i$ and is valid for arbitrary values of $M$. 
We make an interesting observation from \eqref{interference_func4}. The interference measure $|I_{ni}|^2$ depends on the pilot assignment. 
Specifically, $\Psi_m$ denotes the users that are assigned the \textit{m}-th pilot sequence and appears in the denominator of \eqref{interference_func4}. As such, the pilot allocation determines the strength of the LOS interference. We later utilize this observation in our proposed location-aware pilot allocation scheme. Fig.~\ref{int_sin} shows the plot for the interference measure $|I_{ni}|^2$, where $K_n$, $K_i$, $\beta_n$, $\beta_i$, and $p_p$, are fixed as 1 and $M=20$. It is interesting to note that for certain values of the AoA difference between $\textrm{U}_n$ and $\textrm{U}_i$, the value of $|I_{ni}|^2$ is very close to zero.

Despite the fact that \eqref{interference_func4} is derived for ZF linear detector, the extension to other linear detection techniques, such as minimum mean square error (MMSE) and maximum ratio combining (MRC), is straight forward. We highlight that the behavior of $|I_{ni}|^2$, as revealed in Fig.~\ref{int_sin}, will be similar for different type of linear detectors. Against this background, we now detail our proposed location-aware pilot assignment algorithm to improve the uplink sum rate of the massive MIMO network.

\subsection{Location-Aware Pilot Assignment Algorithm}
We now present the step by step procedure for the proposed location-aware pilot assignment algorithm as outlined in \textbf{Algorithm~\ref{algo1}}. The algorithm takes four inputs: $\mathbf{K}=\left[K_1,K_2,\dotsc,K_n\right]$, $\pmb{\theta}=\left[\theta_1,\theta_2,\dotsc,\theta_n\right]$, $\mathbf{d}=\left[d_1,d_2,\dotsc,d_n\right]$, and $\tau$. The algorithm processes the location information and returns the pilot assignment for all the users in the network. We highlight that we perform uplink channel estimation after the location-aware pilot assignment. As such, our proposed algorithm does not require the instantaneous channel state information to improve the sum rate performance. Without loss of generality, we assume the $N = k \times \tau$, where $k$ is an integer. We assume that $k$ is greater than or equal to two because our proposed algorithm focuses on assigning pilot sequences in a pilot contaminated massive MIMO network.

\begin{algorithm}
\caption{Location-aware pilot assignment algorithm} \label{algo1}
\begin{algorithmic}[1]
\Procedure{Pilot Assignment}{$\pmb{\theta}, \mathbf{d},\tau$}
\State $\mathbf{d}' \gets  \textrm{sort}\left(\mathbf{d}, \textrm{ascend} \right)$
\State $N = \textrm{len}\left(\mathbf{d}'\right)$, $padlen = \textrm{rem}\left(N, \tau\right)$
\If{$padlen > 0$}
\State $\mathbf{d}'\gets \left[\mathbf{d}' \hspace{0.2cm} \textrm{zeros}\left(1,padlen\right)\right]$  \Comment{zero padding}
\EndIf
\State $\mathbf{D} \gets \textrm{vec2mat}\left(\mathbf{d}',\tau\right)$;
\State $\mathbf{t}_{1} \gets \textrm{find}\left(\mathbf{d}' == \mathbf{D}\left(1,:\right)\right), \pmb{\theta}_{\mathbf{t}_{1}} \gets \textrm{sin}\left(\pmb{\theta}\left(\mathbf{t}_{1}\right)\right)$
\State $x \gets \textrm{rowlen}\left(\mathbf{D}\right)$
\For{$i\gets 2, x $}
\State $\mathbf{t}' \gets \textrm{find}\left(\mathbf{d}' == \mathbf{D}\left(i,:\right)\right)$ \Comment{\parbox[t]{.28\linewidth}{select the user closer to the BS}}
\State $\pmb{\theta}_{\mathbf{t}'} \gets \textrm{sin}\left(\pmb{\theta}\left(\mathbf{t}'\right)\right)$
\For{$m\gets 1, \textrm{len}\left(\pmb{\theta}_{\mathbf{t}_{1}}\right)$}
\For{$n\gets 1, \textrm{len}\left(\pmb{\theta}_{\mathbf{t}'}\right)$}
\State $d\theta_{mn} \gets \pi(\pmb{\theta}_{\mathbf{t}_{1}}\left(m\right) - \pmb{\theta}_{\mathbf{t}'}\left(n\right))$
\State $\mathbf{I}\left(m,n\right) = |I_{mn}|^2$ \Comment{using eq \eqref{interference_func4}}
\EndFor
\State $\mathbf{\Psi}_{i}\left(n\right) \gets \textrm{min}\left(\mathbf{I}\left(m,:\right)\right)$ \Comment{\parbox[t]{.24\linewidth}{ensure unique assignment in the tier $i$}}
\EndFor
\EndFor
\State $\mathbf{T} = \left[\mathbf{\Psi}_{1},\mathbf{\Psi}_{2},\dotsc,\mathbf{\Psi}_{\tau}\right]^{T}$ \Comment{each column of $\mathbf{T}$ contains the index of the users that are assigned the same pilot sequence}
\EndProcedure
\end{algorithmic}
\end{algorithm}

\begin{description}[leftmargin=*]
\item [Step 1:] First, the proposed algorithm divided the cell area into $k$ tiers of radius $r_1,r_2,\dotsc, r_n$ each containing $\tau$ users. Afterwards, the proposed location-aware pilot assignment algorithm assigns orthogonal pilot sequences to the $\tau$ users in the first tier of the network. The rationale behind orthogonal pilot assignment to the $\tau$ nearest users to the BS is to reduce the pilot contamination. During the uplink training stage, only the NLOS component of the channel is estimated as evident from \eqref{uplink_tx1} and \eqref{chan_est}. The interference power from the NLOS component of $\textrm{U}_n$ depends on the large-scale channel coefficient, i.e, $\beta_n$, which depends on $d_n$. As such, if two users close to the BS are assigned the same pilot sequence, the received signal power at the BS is large, which results in an increased pilot contamination between the users with non-orthogonal pilot sequences.

\item [Step 2:] This step focuses on minimising the interference during the uplink data transmission phase. During the uplink transmission phase, both LOS and NLOS components of the received signal contributes to the interference, which can be confirmed from \eqref{rec_sig} and \eqref{rec_sig_n}. In this phase, the algorithm computes the interference measure $|I_{ni}|^2$ between $\textrm{U}_n$ in the first tier and $\tau$ users in the second tier. We pay special attention to make sure that the AoA are strickly non overlapping, i.e., $d\theta_{ni}\neq0$. It is evident from Fig.~\ref{int_sin} that the interference $|I_{ni}|^2$ is minimum for certain values of $\theta_i$. Unfortunately, due to random nature of the user location, a user may not be located at a place such that $|I_{ni}|^2\approx0$. Next, the algorithm assign the same pilot sequence as the $\textrm{U}_n$ in the first tier to $\textrm{U}_i$ in the second tier with minimum $|I_{ni}|^2$. We clarify that our algorithm follows an aggressive approach, i.e., it assigns pilot sequences to the users closer to the BS first. The benefit of the same pilot assignment to $\textrm{U}_n$ and $\textrm{U}_i$ with minimum $|I_{ni}|^2$ can be observed from \eqref{interference_func4}. We note that the term $\beta_j$ in the denominator of \eqref{interference_func4} is distance-dependent. Assigning same pilot to $\textrm{U}_n$ and $\textrm{U}_i$ with minimum $|I_{ni}|^2$ ensures large spacial separation between the two users, which results in reduction of pilot contamination. This process is repeated for the remaining users in the second tier. As such, we identify the users that should be allocated the same pilot sequences to mitigate pilot contamination. 

\item [Step 3:] Finally, Step 2 is repeated for remaining $n - 2$ tiers. After this step, \textbf{Algorithm~\ref{algo1}} returns a matrix $\mathbf{T}$, where \textit{m}-th column of $\mathbf{T}$ identifies the users that should be assigned the \textit{m}-th pilot sequence, i.e, $\mathbf{\Psi}_{m}$.

\end{description}
We note that \textbf{Algorithm~\ref{algo1}} requires $\left(n-1\right)\left(\tau-1\right)^2$ computation of the interference measure $|I_{ni}|^2$, which is significantly smaller than the $N^2$ computations required for a brute force computation of the interference measure.

\subsection{Performance Metric}
We define a performance metric in accordance with the design of \textbf{Algorithm~\ref{algo1}} as the sum of the interference from the LOS component of the received signal transmission in the uplink. Accordingly, we propose a performance metric as
\begin{align}\label{interference_metric}
  {I}_{\textrm{tot}} =& \sum_{m=1}^{\tau}\sum_{i \in \Psi_{m}}|I_{mi}|^2.
\end{align}

We highlight that it is desirable to have small value of ${I}_{\textrm{tot}}$ in the network. 
\section{Numerical Results}
In this section, we demonstrate the advantages of the proposed pilot assignment scheme through comparing it with the random pilot assignment scheme. In literature, random pilot assignment is widely used for pilot assignment in massive MIMO \cite{Yin2013,Muppirisetty2015,Choi2015,Saxena2015}. Comparison with random assignment is particularly useful because it averages out the the irregularities due to random user locations. Throughout this section, we assume $N=20$, $T=196$ symbols, $\tau=10$, $v=3.8$, and $r_h = 1000$ m. Moreover, we assume that the $r_n$ and $\theta_n$ follow a uniform distribution, where $r_n \sim \left[100~\textrm{m},1000~\textrm{m}\right]$ and $\theta_n \sim \left[0,2\pi\right]$.

\begin{figure}
  \centering
  \includegraphics[width=18pc]{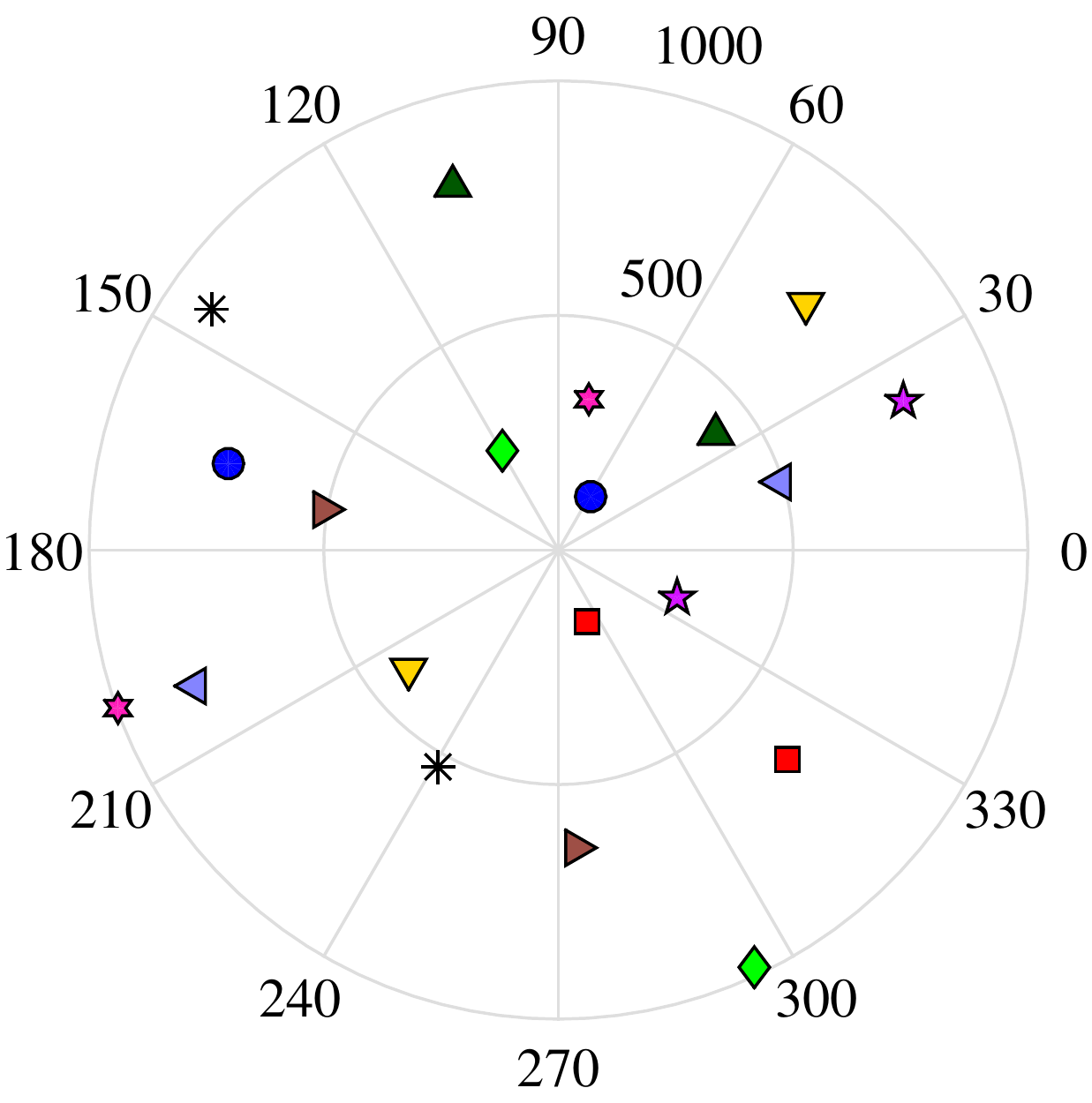}\\
  \caption{Pilot assignment returned by the proposed location-aware pilot assignment algorithm, where the users that are assigned the same pilot sequence are represented by the same shape and color. The BS is located in the center of the cell.} \label{alloc_pairs}
\end{figure}

Fig.~\ref{alloc_pairs} shows the pilot assignment obtained by using the proposed location-aware pilot assignment scheme for given user locations. In Fig.~\ref{alloc_pairs}, the users that are assigned the same pilot sequence are represented by the same shape and color. It is clearly observed that the algorithm assigns the same pilot to the users with large spatial separation, which is consistent with the design of the proposed algorithm.

\begin{figure}
  \centering
  \includegraphics[width=19.1pc]{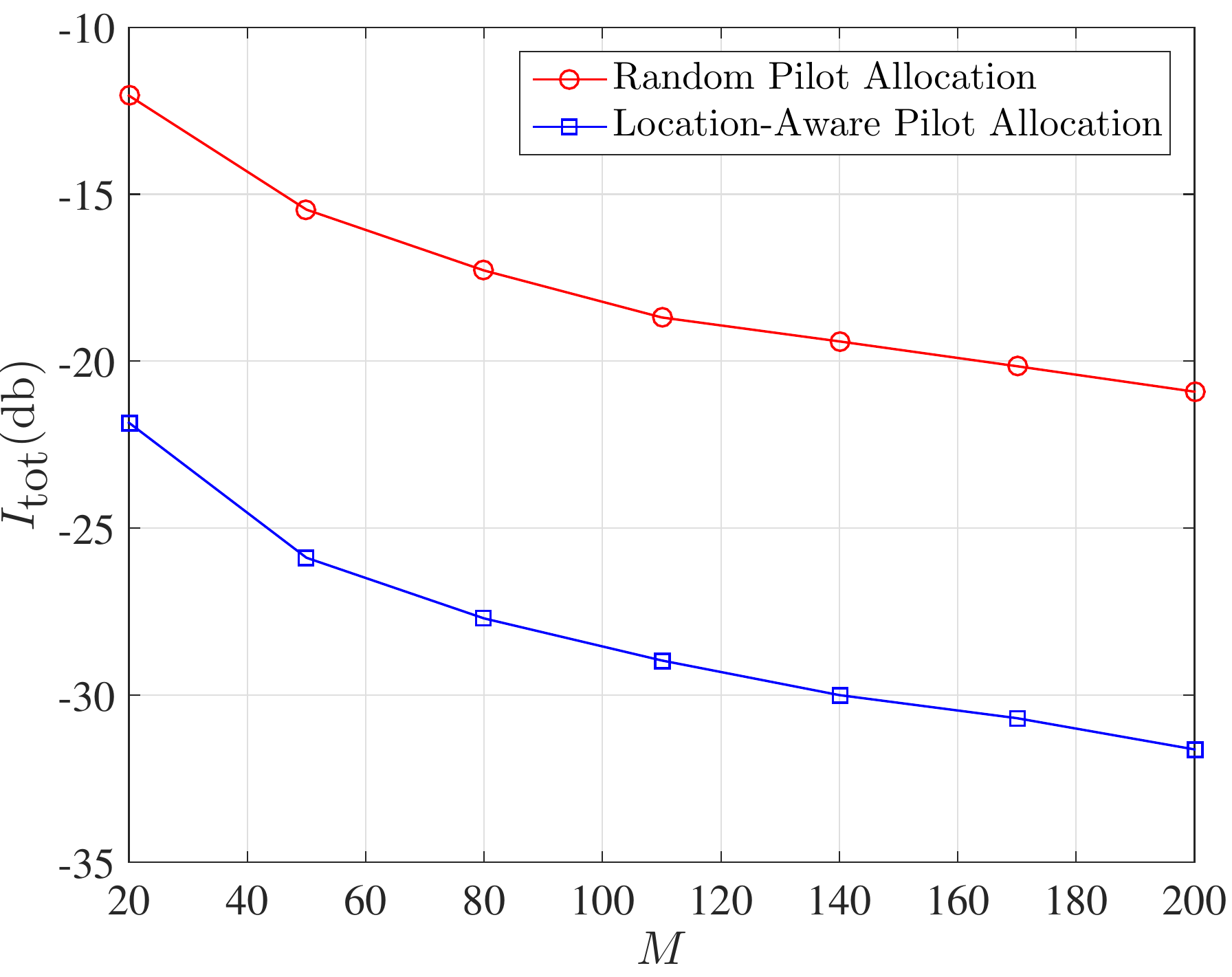}\\
  \caption{Number of antenna at the BS versus the interference measure.}\label{int_fig}
\end{figure}

We now evaluate the performance of the pilot assignment schemes with respect to the proposed performance metric ${I}_{\textrm{tot}}$ given by \eqref{interference_metric}. In this simulation, we assume $K=3$ for all the users in the network. Fig.~\ref{int_fig} shows ${I}_{\textrm{tot}}$ for an average of 200,000 random locations for $N$ users. It can be clearly seen that our proposed location-aware pilot assignment scheme offers less LOS interference as compared to the random pilot assignment because ${I}_{\textrm{tot}}$ is smaller for the proposed scheme. Moreover, ${I}_{\textrm{tot}}$ decreases when $M$ is increased. For example, when $M$ increases from $20$ to $200$, ${I}_{\textrm{tot}}$ decreases from $-12~\textrm{(dB)}$ to $-21.6~\textrm{(dB)}$ for the proposed scheme. Similar behaviour is observed for random pilot assignment. Nevertheless, our proposed pilot assignment scheme continue to offers less LOS interference than the random pilot assignment regardless of the value of $M$.  As such, it is expected that the uplink sum rate of the proposed location-aware pilot assignment scheme will be higher than the random pilot assignment scheme.

\begin{figure}
  \centering
  \includegraphics[width=19.1pc]{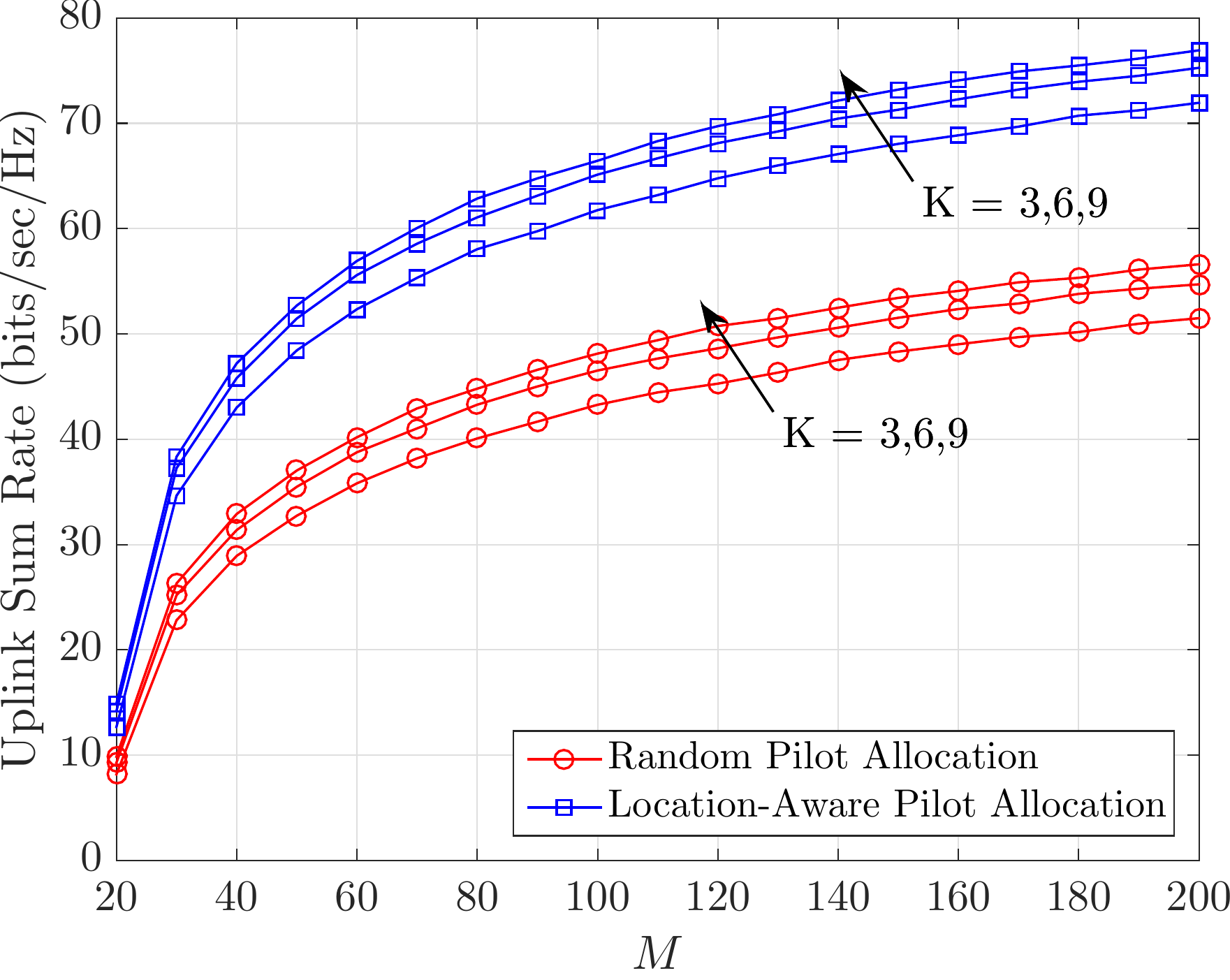}\\
  \caption{Number of antenna at the BS versus uplink sum rate.}\label{sum_rate_fig}
\end{figure}

We next examine the impact of the proposed location-aware pilot assignment scheme on the uplink sum rate of the massive MIMO network. Furthermore, we assume that the \textit{K}-factors for all the users are same. Fig.~\ref{sum_rate_fig} depicts the uplink sum rate for an average over $10,000$ simulations. The uplink sum rate is simulated using \eqref{uplink_sinr_n} and \eqref{sum_rate} for random pilot assignment and the proposed location-aware pilot assignment scheme, respectively. The advantage of the proposed location-aware pilot assignment scheme is clearly visible from Fig.~\ref{sum_rate_fig}. For the given system parameters, location-aware pilot assignment scheme provides up to $20\%$ improvement in uplink sum rate compared to random pilot assignment when $M$ is large. We also note that the uplink sum rate increases with an increase in $K$-factor, which is due to the decrease in interference and the robustness of channel estimation as $K$-factor increases.

\section{Conclusion}
In this work, we proposed a location-aware pilot assignment scheme based the location information of the users in the massive MIMO networks in order to mitigate pilot contamination. Our proposed pilot assignment scheme calculates the interference offered from the LOS component of the received signal and allocates the same pilot sequences to the users with minimum LOS interference. Comparison with the random pilot assignment scheme demonstrates the advantages offered by the proposed scheme in term of achieving a higher uplink sum rate. Our work provides practical insights to better understand the pilot contamination in Rician fading channels and how location information can be utilized to reduce the impact of pilot contamination.


\end{document}